\documentstyle[12pt,epsfig]{article}
\newcommand{\bea}{\begin{eqnarray}}
\newcommand{\eea}{\end{eqnarray}}
\newcommand{\beq}{\begin{equation}}
\newcommand{\eeq}{\end{equation}}
\newcommand{\bay}{\begin{array}}
\newcommand{\eay}{\end{array}}
\def \bo{B^0}
\def \ob{\overline{B}^0}
\def \ok{\overline{K}^0}
\def \s{\sqrt{2}}
\def \ob{\overline{B}^0}
\def \cn{Collaboration}
\def \ite{{\it et al.}}

\begin{document}

\begin{center}
\large
{\bf PENGUINS IN $B$ DECAYS~\footnote{Invited talk at the Ninth International 
Symposium on Heavy Flavor Physics, Caltech, Pasadena, CA, Sept. 10$-$13, 2001.}}
\end{center}

\bigskip

\begin{center}
\normalsize
Michael Gronau\\

\medskip
{\it Department of Physics, Technion-Israel Institute of Technology \\
Technion City, 32000 Haifa, Israel}

\bigskip
{\bf Abstract}

\end{center}

\begin{quote}
We report on recent progress in studying two aspects of $B$ physics, in which 
penguin amplitudes play an important role: 
\begin{enumerate}
\item Bounds on {\em penguin pollution} in $B^0(t) \to \pi^+\pi^-$ constraining the CKM 
parameters $\rho$ and $\eta$, and a lower bound on $B^0 \to \pi^0\pi^0$ improving 
precision in $\sin 2\alpha$.
\item A suggestion for measuring the photon polarization in {\em electroweak 
penguin} decay, $B \to K_1(1400)\gamma$, providing a test of the 
Standard Model and a probe for new physics.
\end{enumerate}
\end{quote}

\bigskip

\section{Introduction}
When being asked to choose a topic for my talk at this conference, I responded 
without hesitation by making the above choice. The topic of penguins in $B$ decays 
is quite broad and covers a large variety of aspects, some of which are discussed by 
other speakers at this conference. The two particular aspects to which I will address 
my talk are almost as old as the entire field of $B$ physics. Let me remind you how 
penguins entered heavy flavor physics.
{\em Gluonic penguin diagrams} were introduced twenty five years ago \cite{Shifman} 
when analyzing QCD effects in hadronic $K$ meson decays. Shortly afterwards penguin 
amplitudes were shown to play an important role in direct CP violation, first studied 
in $K$ decays \cite{GW} and soon afterwards in $B$ decays \cite{BSS}. A couple of 
years later it was realized that intermediate heavy fermions, such as a heavy top 
quark, imply sizable {\em electroweak penguin amplitudes} governing radiative $K$ 
\cite{Inami} and $B$ \cite{CO} decays. These historical remarks lead naturally to 
my two topics.

\subsection{Penguin pollution in $B^0(t)\to\pi^+\pi^-$}
Calculations of direct CP violation in $B$ decays, due to interference between tree 
and penguin amplitudes, involve theoretical uncertainties in nonperturbative 
hadronic 
matrix elements of weak operators and uncertainties in final state interaction phases 
\cite{Neubert}. This poses a difficulty in interpreting a measurement of the 
time-dependent CP asymmetry in $B^0(t)\to\pi^+\pi^-$ in terms of a fundamental CKM 
phase \cite{MG, LP}. This problem, known as the problem of {\em penguin pollution}, 
is dealt with in Section 2. I will show first that a crude asymmetry measurement and 
an approximate knowledge of the ratio of penguin to tree amplitudes improve 
significantly our present knowledge of CKM parameters. 

I will then discuss the cleanest way of resolving the penguin pollution \cite{GL}, 
which is based on applying isospin symmetry to the system of all three decays 
$B^0\to\pi^+\pi^-, B^+\to\pi^+\pi^0$ and $B^0\to \pi^0\pi^0$. The most challenging 
experimental task in this method is measuring decay rates into two neutral 
pions while distinguishing between $B^0$ and $\bar B^0$ decays. I will 
report on recent theoretical progress made in order to overcome this difficulty. 

\subsection{Photon polarization in radiative $B$ decays}
In the Standard Model radiative $B$ decays have one unambiguous signature which has 
not yet been tested. Namely, in decays of $B^-$ and $\bar B^0$ (containing a $b$ 
quark) the emitted photon is left-handed polarized, while in $B^+$ and $B^0$ it is 
right-handed. 
This prediction of {\em maximal parity violation}, holds to within a percent and, 
in principle, can serve as a {\em precision test} of the Standard Model. 
Deviations from this prediction are sensitive probes of new physics. In Section 3 I 
will survey several suggestions for studying photon helicity effects in radiative $B$ 
and $\Lambda_b$ decays, focusing on a particular method. A measurement of the photon 
polarization in $B\to K\pi\pi\gamma,~m(K\pi\pi) = 1400$ MeV, through decay 
particle angular distributions, will be shown to be feasible at currently operating 
$B$ factories. 

\medskip
Finally, Section 4 contains several concluding remarks.

\section{Bounds on penguin pollution in $B^0\to \pi^+\pi^-$}
\subsection{CP asymmetry in $B^0(t)\to\pi^+\pi^-$}
The weak phase $\alpha \equiv {\rm arg}(-V^*_{tb}V_{td}/V^*_{ub}V_{ud}) = \pi-\beta-
\gamma$ occurs in the 
time-dependent rate of $B^0(t)\to \pi^+\pi^-$ and would dominate its asymmetry
if only a {\em tree} amplitude $T$ contributed. In reality this process involves a 
second amplitude $P$ due to {\em penguin} operators which carries a different weak
phase than the dominant tree amplitude, 
\beq
A(B^0 \to \pi^+ \pi^-) = |T|e^{i \delta_T} e^{i \gamma} +
 |P| e^{i \delta_P}~~~.
\eeq
The two terms contain CKM factors $V^*_{ub}V_{ud},~V^*_{cb}V_{cd}$ and weak 
phases $\gamma$ and $0$, respectively.
This leads to a generalized form of the time-dependent asymmetry,
which includes in addition to the $\sin(\Delta mt)$ term a $\cos(\Delta mt)$
term due to direct CP violation  \cite{MG}
\beq\label{asymmet}
{\cal A}(t) = C_{\pi \pi}\cos(\Delta mt) + S_{\pi \pi}\sin(\Delta mt)~~,
\eeq
\beq\label{CSpipi}
C_{\pi \pi} \equiv a_{\rm dir} = \frac{1 - |\lambda_{\pi \pi}|^2}{1 +
|\lambda_{\pi \pi}|^2}~~~, 
S_{\pi \pi} \equiv \sqrt{1-a^2_{\rm dir}}\sin 2(\alpha + \Delta\alpha) = \frac{2 
{\rm Im}(\lambda_{\pi \pi})}{1 + |\lambda_{\pi \pi}|^2}~~, 
\eeq
where
\beq
\lambda_{\pi \pi} \equiv e^{-2i \beta} \frac{A(\ob \to \pi^+ \pi^-)}
{A(B^0 \to \pi^+ \pi^-)}~~~.
\eeq
In the absence of the penguin amplitude one would have $C_{\pi\pi} = \Delta\alpha = 0, 
S_{\pi\pi}= \sin 2\alpha$.
\begin{figure}[t]
\centerline{\epsfysize = 3 in \epsffile{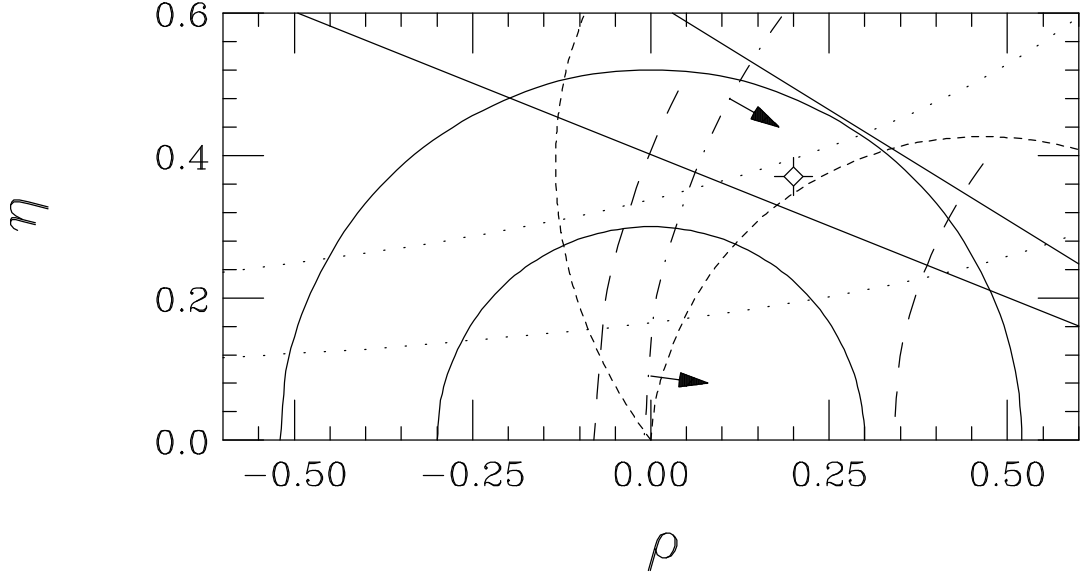}}
\caption{Constraints on parameters of the CKM matrix.  Solid circles denote
limits on $|V_{ub}/V_{cb}| = 0.090 \pm 0.025$ from charmless $b$ decays.
Dashed arcs denote limits from $\bo$--$\ob$ mixing.  Dot-dashed arc denotes
limit from $B_s$--$\overline{B}_s$ mixing.  Dotted hyperbolae are associated
with limits on CP-violating $K^0$--$\ok$ mixing.
Limits of $\pm 1 \sigma$ from CP asymmetries in $B^0 \to J/\psi K_S$,
$\sin(2 \beta) = 0.79 \pm 0.10$, are shown by the solid rays.
The small dashed lines represent constraints due to $1\sigma$ bounds 
$-0.53 \le S_{\pi\pi} \le 0.59$,
with $0.21 \le |P/T| \le 0.34$. The plotted point lies in the middle of the allowed 
region.
\label{fig:re}}
\end{figure} 
The time-dependent asymmetry measurement provides two equations for $C_{\pi\pi}$ and 
$S_{\pi\pi}$ in terms of $|P/T|,~\delta \equiv \delta_P - \delta_T$ and $\alpha$. 
This is insufficient for a determination of $\alpha$.
Knowledge of $|P/T|$ would, in principle, enable this determination up to discrete 
ambiguities. A crude estimate \cite{DGR}, $|P/T| = 0.3 \pm 0.1$, was obtained 
several years ago by applying flavor SU(3) 
to the ratio of $B\to\pi\pi$ and $B\to K\pi$ branching ratios first 
measured by CLEO \cite{pipi}. A more precise evaluation including SU(3) 
breaking, from averaging recent CLEO, Belle and BaBar branching ratios 
\cite{CBB}, yields \cite{Asym} $|P/T| = 0.276 \pm 
0.064$. In a QCD factorization approach, where absolute hadronic weak amplitudes 
including strong phases are calculated, one finds \cite{BBNS} $|P/T| = 0.285 \pm 0.076$.

The BaBar Collaboration reported recently the first measurements of $C_{\pi\pi}$ and 
$S_{\pi\pi}$ \cite{BaBasy}
\beq
C_{\pi \pi} = -0.25^{+0.45}_{-0.47} \pm 0.14~~,~~~
S_{\pi \pi} = 0.03^{+0.53}_{-0.56} \pm 0.11~~.
\eeq
This asymmetry measurement is still very crude. Nonetheless, to anticipate the 
significance of future improvements, we have studied recently \cite{Asym} the 
implication of present $1\sigma$ bounds, $-0.53 \le S_{\pi\pi} \le 0.59$, on CKM 
parameters. Assuming that $\delta$ is small \cite{Neubert, BBNS}, one has 
\beq
S_{\pi \pi} \simeq \sin[2(\alpha + \Delta\alpha)]~~,~~~ 
\tan \alpha = \frac{\eta}{\eta^2 - \rho(1-\rho)}~~,~~~ \tan{\Delta \alpha}
= \frac{\eta|P/T|}{\sqrt{\rho^2 + \eta^2} + \rho|P/T|}~~~.
\eeq
Using $0.21 \le |P/T| \le 0.34$ we found in \cite{Asym} that {\em the above 
$1\sigma~S_{\pi \pi}$ bounds exclude more than half of the $(\rho,\eta)$ parameter 
space \cite{PDG} allowed by all other constraints on CKM parameters.} 
Results are shown in Fig.~1.
The exclusion plot due to $S_{\pi\pi}$ is rather striking in view of the large 
uncertainties assumed here for 
$S_{\pi\pi}$ and $|P/T|$ which imply an uncertainty $\Delta\alpha$ in $\alpha$ as 
large as about $\pm 20^\circ$. The assumption of a small $\delta$ can be tested by 
improving limits on $C_{\pi\pi}$.

\subsection{Combining $B^0\to\pi^+\pi^-$ with $B^+\to\pi^+\pi^0,~B^0\to\pi^0\pi^0$}
The isospin method \cite{GL} requires measuring also the 
time-integrated rates of $B^+\to\pi^+\pi^0$, $B^0\to\pi^0\pi^0$
and their charge-conjugates. The three $B\to\pi\pi$ amplitudes 
obey an isospin triangle relation,
\beq\label{iso}
A(B^0\to\pi^+\pi^-)/\s + A(B^0\to\pi^0\pi^0) = A(B^+\to \pi^+\pi^0)~~,
\eeq
while a similar relation holds for the charge-conjugate processes.
One uses the different isospin properties of the penguin ($\Delta I=1/2$) and tree 
($\Delta I=1/2, 3/2$) contributions and the well-defined weak phase ($\gamma$) of the 
tree amplitude. The electroweak penguin amplitude is very small \cite{DH}, and can be 
dealt with as function as $V_{td}/V_{ub}$ in the isospin symmetry limit \cite{GPY}.
Laying the two isospin triangles such that they have a common side, 
$A(B^+ \to\pi^+\pi^0) = A(B^-\to\pi^-\pi^0)$, the angle between $A(B^0\to\pi^+\pi^-)$ 
and $A(\bar B^0\to\pi^+\pi^-)$ is $2\Delta\alpha$ which then determines $\alpha$ from 
the asymmetry (\ref{asymmet}).

While this method is the cleanest theoretically, it suffers from the experimental 
difficulty associated with the two neutral pion mode. In fact, this introduces three 
kinds of practical complications:
\begin{enumerate}
\item $B^0\to \pi^0\pi^0$ is argued to be {\em color-suppressed} and, although there 
exists no reliable calculation for this branching ratio, it is customarily assumed 
to be much smaller than the other two branching ratios.
\item One requires neutral $B$ {\em flavor tagging} in order to distinguish between 
$B^0\to\pi^0\pi^0$ and $\bar B^0\to\pi^0\pi^0$.
\item Neutral pions have a somewhat {\em lower detection efficiency} than charged 
pions.
\end{enumerate}

In the subsequent discussion we will show how to overcome the first two obstacles,
which lead one to ask the following question: Assuming that 
one has only an upper bound on the sum of $B^0$ and $\bar B^0$ decay 
branching ratios to $\pi^0\pi^0$, can one put an upper limit on $|\Delta\alpha|$?
This question was addressed a few years ago, and a partial answer was 
given under the assumption that the sum of rates of $B^+\to\pi^+\pi^0$ and its
charge conjugate is known. An upper bound, based on right angle isospin triangles, 
was given in terms of the ratio of charged-to-neutral $B$ lifetimes \cite{Blifes}
and the ratio of charge-averaged branching ratios ${\cal B}(B\to \pi^0\pi^0)/
{\cal B}(B^\pm\to \pi^{\pm}\pi^0)$ \cite{GQ}:
\beq\label{GRQ}
|\sin(\Delta\alpha)| \le
\sqrt{\frac{r_{\tau}{\cal B}(B\to \pi^0\pi^0)}{{\cal B}(B^\pm\to \pi^{\pm}\pi^0)}}~~,
~~~r_{\tau}\equiv \frac{\tau_{B^+}}{\tau_{B^0}} = 1.068 \pm 0.016~~.
\eeq
A slight improvement, involving the direct CP asymmetry in $B^0\to\pi^+\pi^-$, 
$a_{\rm dir}$, as well 
as an independent bound assuming the knowledge of ${\cal B}(B\to \pi^+\pi^-)$ instead 
of ${\cal B}(B^{\pm}\to \pi^{\pm} \pi^0)$, were suggested in \cite{Charles}.

Although these two bounds are somehow related to the isospin triangles, neither of 
them involves all three $B\to\pi\pi$ processes, implying that the saturation of these
bounds may be inconsistent 
with the closure of the triangles. Thus, the real question is what is the maximum value
of $|\Delta\alpha|$, consistent with the closure, for given ${\cal B}(B\to \pi^+\pi^-)$ 
and ${\cal B}(B^{\pm}\to \pi^{\pm}\pi^0)$ and for an upper bound on ${\cal B}(B\to
\pi^0\pi^0)$. The correct answer to this question was found recently \cite{GLSS},
\beq\label{BOUND}
\cos (2\Delta\alpha) \ge \frac{(B^{+-}/2 - B^{00} + B^{+0}/r_{\tau})^2 - 
B^{+-}B^{+0}/r_{\tau}}{\sqrt{1-a^2_{\rm dir}}B^{+-}B^{+0}/r_{\tau}}~~,
\eeq
where $B^{ij}$ are corresponding charge-averaged branching ratios. This bound is 
stronger than Eq.~(\ref{GRQ}) and the bound \cite{Charles}, as demonstrated in 
\cite{GLSS}.
A crucial difference between Eq.~(\ref{BOUND}) and the earlier bounds is that 
(\ref{BOUND}) includes also a lower bound on $B^{00}$, following from the triangle 
construction:
\beq\label{B00}
B^{00} \ge B^{+0}/r_{\tau} + B^{+-}/2 -\sqrt{(1 + \sqrt{1-a^2_{\rm dir}})
B^{+-}B^{+0}/r_{\tau}}
\ge (\sqrt{B^{+0}/r_{\tau}} - \sqrt{B^{+-}/2})^2~~.
\eeq

The advantage of the two bounds Eqs.~(\ref{BOUND}) and (\ref{B00}) over (\ref{GRQ})
and \cite{Charles} was demonstrated in \cite{GLSS} when using the present world
averaged branching ratios in units of $10^{-6}$ \cite{CBB}
\beq
B^{+-} = 4.4 \pm 0.9~,~~~B^{+0} = 5.6 \pm 1.5~,~~~
B^{00} < 5.7 ~(90\% ~{\rm C.L.})~.
\eeq
In order to illustrate the future potential power of these bounds in reducing the 
error in $\alpha$ due to penguin pollution, we list below values of the three 
branching ratios with corresponding errors, which were measured and which can be 
measured at $B$ factories with higher integrated luminosities. Errors in $B^{ij}$  
scale down as $1/\sqrt{\rm luminosity}$. For illustration purpose, we will take the 
future central value of $B^{+0}$ to be less than $1\sigma$ above its present central 
value.
\begin{center}
\begin{tabular}{c c c c} \hline
\qquad luminosity: & $30~{\rm fb}^{-1}$  & $120~{\rm fb}^{-1}$ & $500~{\rm fb}^{-1}$ 
\\ \hline
$B^{+-}$ & $4.4 \pm 0.9$ & $4.4 \pm 0.4$ & $4.4 \pm 0.2$ \\
$B^{+0}$ & $5.6 \pm 1.5$  & $7.0 \pm 0.8$ & $7.0 \pm 0.4$ \\
$B^{00}$ & $< 5.7$ & $< 1.4$ & $< 0.4$ or seen  \\ \hline
$B^{00}$ & $\ge 0.78 \pm 0.62$ & $\ge 1.35 \pm 0.38$ & $\ge 1.35 \pm 0.19$ \\ \hline
\end{tabular}
\end{center}
The last line in the table gives the lower bounds on $B^{00}$ obtained from 
Eq.~(\ref{B00}) for the corresponding values of $B^{+-}$ and $B^{+0}$. 

Thus, while an upper bound on $B^{00}$ can be obtained from a direct measurement,
useful lower bounds follow from measuring the other two branching ratios. 
If $B^{00}$ is not very small, which does not seem unlikely in view of the present
values of $B^{+-}$ and $B^{+0}$, one may be able
to restrict its values from above and below to a narrow range. Consequently, the 
uncertainty in measuring $\alpha$ becomes small. Assuming, for instance,
that one finds $1.2 \le B^{00} \le 1.3$, permitted by the lower bound derived for 
$500~{\rm fb}^{-1}$, one obtains from Eq.~(\ref{BOUND}) $|\Delta\alpha| < 9^\circ$.
In comparison, the bound (\ref{GRQ}) implies only $|\Delta\alpha| < 26^\circ$. 
We stress that this demonstration of a rather precise determination of $\alpha$ 
(where the uncertainty follows only from penguin pollution) assumes no separation 
between $B^0$ and $\bar B^0$ decays to $\pi^0\pi^0$. Neutral $B$ flavor tagging will 
reduce the uncertainty further.

\section{The photon polarization in $b\to s\gamma$}
The present agreement between experiment and the Standard Model (SM) prediction for 
the rate of inclusive $B \to X_s\gamma$ is reasonable, at a level of 20$\%$ 
\cite{Bsgamma}. However, one basic feature, the 
left-handedness of the emitted photon in $b \to s\gamma$, has never been tested. 
The photon is predominantly left-handed, since the recoil $s$ quark which couples 
to a $W$ is left-chiral. In several extensions of the SM, including left-right 
symmetric \cite{LRM} and supersymmetric models \cite{SUSY}, in which decay amplitudes 
involve $W_L-W_R$ mixing and scalar exchange, the photon can acquire a large right-handed 
component without affecting substantially the inclusive rate.

Formally, the effective weak Hamiltonian for radiative $b$ decays contains two Wilson 
coefficients, $C_{7L}$ and $C_{7R}$, multiplying operators, ${\cal O}_{7L}$ and 
${\cal O}_{7R}$, describing left and right handed emitted photons,  
\beq
{\cal H}_{\rm rad} = -\frac{4G_F}{\sqrt2} V_{tb} V_{ts}^*\left( C_{7L}{\cal O}_{7L}
+  C_{7R}{\cal O}_{7R}\right)~,~~~{\cal O}_{7L,R} \equiv \frac{e}{16\pi^2} m_b 
\bar s\sigma_{\mu\nu}\frac{1 \pm \gamma_5}{2} bF^{\mu\nu}~.
\eeq
The photon polarization in inclusive $b \to s\gamma$ is 
\beq 
\lambda_{\gamma}\equiv {|C_R|^2 -|C_L|^2\over |C_R|^2 +|C_L|^2}~.
\eeq
In the SM, where $C_{7R}/C_{7L}=m_s/m_b$,  the polarization in exclusive decays is 
$\lambda_\gamma = -1$ within a percent, also when modified by long distance 
hadronic effects \cite{GriP}. This prediction can provide 
precision tests of the SM and sensitive probes for new physics.

Several ways of carrying out such measurements were proposed in the past. They
require very high luminosity $B$ factories or new experimental facilities. We will 
describe very briefly these early suggestions, and will focus our attention on a recent 
proposal which is feasible at currently operating $B$ factories.

\subsection{CP asymmetry in $B^0(t) \to X^{CP}_{s(d)}\gamma$} 
Consider the time-dependent rate of \cite{AGS} $B^0(t) \to X^{CP}_{s(d)}\gamma$, where 
$X^{CP}_s = K^{*0} \to K_S\pi^0$ or $X^{CP}_d = \rho^{0} \to \pi^+\pi^-$. The 
time-dependent CP asymmetry follows from interference between $B^0$ and $\bar B^0$ 
decay amplitudes into a common state of definite photon polarization, and is 
proportional to $C_{7R}/C_{7L}$. For instance, in the SM the asymmetry in $B^0(t) 
\to f,~ f =K^{*0}\gamma\to (K_S\pi^0)\gamma$, is given by
\beq
{\cal A}(t) \equiv \frac{\Gamma(B^0(t) \to f) - \Gamma(\bar B^0(t) \to f)}
{\Gamma(B^0(t) \to f) + \Gamma(\bar B^0(t) \to f)}
= {2A_L A_R \over A_L^2 + A_R^2}\sin 2\beta\sin(\Delta mt)~,
\eeq
where $A_{L(R)}$ is the amplitude for a left (right) handed photon in $\bar B \to
\bar K^*\gamma$. In the SM
one expects $A_R/A_L \leq 0.05$ in the presence of long distance effects, 
whereas in extensions of the SM this ratio may be much larger \cite{AGS}.

\subsection{Angular distribution in $\bar B\to \bar K^*\gamma \to \bar K\pi e^+ e^-$}
Consider the decay distribution in this process as function of the angle $\phi$ 
between the $\bar K\pi$ and $e^+e^-$ planes, where the photon can be virtual 
\cite{Kim} or real, converting in the beam pipe to an electron-positron pair 
\cite{GrPi}. The $e^+e^-$ plane acts as a polarizer, the distribution in $\phi$
is isotropic for purely circular polarization, and the angular distribution is 
sensitive to interference between left and right polarization. One finds
\beq
\frac{d\sigma}{d\phi} \propto 1 + \xi {A_L A_R 
\over A_L^2 + A_R^2}\cos(2\phi + \delta)~,
\eeq
where the parameters $\xi$ and $\delta$ are calculable and involve hadronic physics.
 
\subsection{Forward-backward asymmetry in $\Lambda_b \to \Lambda\gamma \to p\pi\gamma$}
The forward-backward asymmetry of the proton with respect to the $\Lambda_b$ in the 
$\Lambda$ rest-frame is proportional to the photon polarization $\lambda_\gamma$ 
\cite{MaRe}. Using polarized $\Lambda_b$'s from extremely high luminosity 
$e^+e^-$ $Z$ factories, one can also measure the forward-backward asymmetry of the
$\Lambda$ momentum with respect to the $\Lambda_b$ boost axis \cite{HK}. This 
asymmetry is proportional to the product of the $\Lambda_b$ and photon polarizations.

\subsection{Angular distribution in $B\to K_1(1400)\gamma \to K\pi\pi\gamma$}
In order to measure the photon polarization $\lambda_\gamma$ in radiative $B$ decays 
through the recoil hadron distribution, one requires that the hadrons consist of at 
least three particles. 
A hadronic quantity which is proportional to $\lambda_\gamma$ must be 
parity odd. The pseudoscalar quantity, which contains the smallest number 
of hadron momenta, is a triple product.
The idea is then to measure an expectation value $\langle \vec 
p_{\gamma}\cdot (\vec p_1\times \vec p_2)\rangle$, where $\vec p_1$ and $\vec p_2$
are momenta of two of the hadrons. Since the triple product is also time-reversal
odd, a nonzero expectation value requires a phase due to final state interactions.
While in general such a phase would be incalculable, there are special cases where 
the decay occurs through two isospin-related intermediate resonance states, and the 
phase can be calculated simply in terms of Breit-Wigner forms \cite{GGPR}. 

Consider the decays $B^+ \to K^+_1(1400)\gamma$ and $B^0 \to K^0_1(1400)\gamma$, 
where $K^+_1$ and $K^0_1$ are observed through 
\bea
K^+_1(1400)\to \left\{
\begin{array}{c}
 K^{*+}\pi^0 \\
 K^{*0} \pi^+ 
\end{array}
\right\} \to K^0 \pi^+ \pi^0~,~~~
K^0_1(1400)\to \left\{
\begin{array}{c}
 K^{*+}\pi^- \\
 K^{*0} \pi^0 
\end{array}
\right\} \to K^+ \pi^- \pi^0~.
\eea
Two Breit-Wigner amplitudes interfere due to intermediate $K^{*+}$ and $K^{*0},~ 
{\cal B}(K_1\to K^*\pi) = 0.94\pm 0.06$ \cite{PDG}. Decay to $\rho K$ will be 
neglected at this point, ${\cal B}(K_1\to \rho K) = 0.03\pm 0.03$. The two $K^*$ 
amplitudes are related by isospin; therefore phases other 
those related to the Breit-Wigner phase cancel. The decay $K_1 \to K^*\pi$ is dominated 
by an $S$ wave and involves a small $D$ waves, where the $D/S$ ratio of rates is
$|A_D/A_S|^2 = 0.04 \pm 0.01$ \cite{PDG}. Using Lorentz invariance, it is straightforward 
to write down the decay amplitude for $B\to (K\pi\pi)_{K_1}\gamma$, and to calculate
the decay distribution \cite{GGPR},
\beq
\frac{d\Gamma}{ds_{13}ds_{23}d\cos\theta} \propto |\vec J|^2(1 + \cos^2\theta)
+ \lambda_{\gamma} 2{\rm Im}\left (\hat n\cdot (\vec J\times\vec J^*)\right )
\cos\theta~,
\eeq
where
\bea
\vec J = & &\vec p_1\left [\left ((1 - \frac{m_K^2-m_{\pi}^2}{m_{K^*}^2}) 
(1 - \kappa(p_{K_1}\cdot p_1 - m^2_{\pi})) - 2\kappa p_1\cdot p_2 \right )
B(s_{23}) - 2B(s_{13})\right ] \nonumber\\
& & - (p_1 \leftrightarrow p_2)~,
\eea
\beq
B(s) = \left( s - m_{K^*}^2 - im_{K^*} \Gamma_{K^*}\right)^{-1}~, 
~~~s_{ij}=(p_i + p_j)^2~.
\eeq
The parameter 
$\kappa = [0.38 + 8.66|A_D/A_S|e^{i(\delta_D-\delta_S)}]
[1 + 0.71|A_D/A_S|e^{i(\delta_D-\delta_S)}]^{-1}{\rm GeV}^{-2}$, $\delta_D-\delta_S = 
(260 \pm 20)^{\circ}$, parametrizes the $D$ wave contribution \cite{PDG}.
$p_1$ and $p_2$ are the two pion momenta, $p_3$ is the $K$ momentum, and $\theta$ is 
the angle between the 
normal to the decay plane $\hat n\equiv (\vec p_1 \times \vec p_2)/|\vec p_1 \times 
\vec p_2|$ and $-\vec p_\gamma$, all measured in the $K_1$ rest frame. A useful
definition of the normal is in terms of the slow and fast pion momenta, $(\vec p_{\rm 
slow}\times \vec p_{\rm fast})/|\vec p_{\rm slow}\times \vec p_{\rm fast}|$. The 
angle between this norml and $-\vec p_\gamma$ will be denoted by $\tilde\theta$.

The decay distribution exhibits an up-down asymmetry of the photon momentum with 
respect to the $K_1$ decay plane. The up-down asymmetry is proportional to the
photon polarization. When integrating over the entire Dalitz plot one finds 
\beq
{\cal A}_{\rm up-down} \equiv \frac{\int_0^{\pi/2}\frac{d\Gamma}{d\cos\tilde\theta}
d\cos\tilde\theta -  \int_{\pi/2}^\pi \frac{d\Gamma}{d\cos\tilde\theta}d\cos\tilde\theta}
{\int_0^\pi \frac{d\Gamma}{d\cos\tilde\theta}d\cos\tilde\theta} = (0.34 \pm 0.05)
\lambda_\gamma~.
\eeq
The uncertainty follows from experimental errors in the $\rho K$ and in the $D$ wave 
amplitudes.
In the SM, where $\lambda_\gamma \approx -1$, the asymmetry is $34 \pm 5\%$ and the 
polarization signature is unambiguous: {\em In $B^-$ and $\bar B^0$ decays the photon 
prefers to be emitted in the hemisphere of 
$\vec p_{\rm slow}\times \vec p_{\rm fast}$, while in $B^+$ and $B^0$ it is more 
likely to be emitted in the opposite hemisphere}.

Is this measurement feasible at currently operating $B$ factories?
A $3\sigma$ measurement of a $34\%$ up-down asymmetry requires about 80 
reconstructed $B\to K_1(1400)\gamma \to K\pi\pi\gamma$ events, including 
charged and neutral $B$ and $\bar B$ decays. Assuming ${\cal B}(B\to
K_1\gamma) = 0.7\times 10^{-5}$ \cite{BR} and including $K_1$ and $K^*$ 
branching ratios to the relevant charge states, one finds that this number of
reconstructed events can be obtained from a total of $2\times 10^7$ $B\bar B$ pairs, 
including charged and neutrals. This number has already been produced at $e^+e^-$
colliders. Since we ignored experimental efficiencies, resolution and background, 
one may have to wait a year or so before obtaining the required number of events.

The region of $K\pi\pi$ invariant mass around $1400$ MeV contains also two other 
resonances, a spin 2 positive-parity $K^*_2(1430)$ which has already been observed in 
radiative $B$ decays \cite{CLEO, Belle}, and a vector state $K^*_1(1410)$, both of 
which decay to $K^*\pi$. (A nonresonant contribution in a narrow bin around
$m(K\pi\pi)=1400$ MeV is expected to be very small.) The $K^*_2$ decays involve a 
smaller up-down asymmetry with 
the same sign as $K_1$ \cite{GGPR} (although the integrated asymmetry vanishes), 
while the decay of $K^*_1$ 
is up-down symmetric. Whereas the overall polarization signature is unchanged, the
integrated up-down asymmetry would be diluted relative to the asymmetry from 
$K_1(1400)$ if all three resonance contributions would be added. It is therefore 
useful to isolate the $K_1$ from the other two resonances. This 
can be achieved by applying to the data an angular decay distribution characterizing 
an axial vector particle.  

\section{Concluding remarks}
\begin{itemize}
\item While the study of CP asymmetry in $B^0\to\pi^+\pi^-$  in terms of 
$\sin 2\alpha$ is complicated by a penguin amplitude, even crude limits on 
the asymmetry may exclude a large part of the presently allowed CKM parameter space.
\item A lower bound on the charge-averaged branching ratio of $B\to\pi^0\pi^0$ from 
measured $B\to\pi^+\pi^-$ and $B^{\pm}\to \pi^{\pm}\pi^0$ may reduce 
the uncertainty of measuring $\sin 2\alpha$, without carrying out the complete 
isospin analysis.
\item The photon polarization in $b \to s\gamma$, predicted to be left-handed in the
Standard Model, can be measured through angular 
decay distributions in $B \to K\pi\pi\gamma$ around $m(K\pi\pi)=1400$ MeV. 
\end{itemize}
I expect that in a year these measurements will lead to interesting and useful results.

\section{Acknowledgments}
I am grateful to D. Atwood, Y. Grossman, D. London, D. Pirjol, J. L. Rosner, A. Ryd, 
N. Sinha, R. Sinha and A. Soni for enjoyable collaborations on work discussed in this talk.
I wish to thank SLAC and the Aspen Center for Physics where this talk was prepared. 
This work was supported in part by the Fund for the Promotion of Research at the 
Technion, by the Israel Science Foundation founded by the Israel Academy of Sciences 
and Humanities, and by the U. S. -- Israel Binational Science Foundation through 
Grant No.\ 98-00237.

\def \ajp#1#2#3{Am.\ J. Phys.\ {\bf#1}, #2 (#3)}
\def \apny#1#2#3{Ann.\ Phys.\ (N.Y.) {\bf#1}, #2 (#3)}
\def \app#1#2#3{Acta Phys.\ Polonica {\bf#1}, #2 (#3)}
\def \arnps#1#2#3{Ann.\ Rev.\ Nucl.\ Part.\ Sci.\ {\bf#1}, #2 (#3)}
\def \art{and references therein}
\def \cmts#1#2#3{Comments on Nucl.\ Part.\ Phys.\ {\bf#1}, #2 (#3)}
\def \cn{Collaboration}
\def \cp89{{\it CP Violation,} edited by C. Jarlskog (World Scientific,
Singapore, 1989)}
\def \efi{Enrico Fermi Institute Report No.\ }
\def \epjc#1#2#3{Eur.\ Phys.\ J. C {\bf#1}, #2 (#3)}
\def \f79{{\it Proceedings of the 1979 International Symposium on Lepton and
Photon Interactions at High Energies,} Fermilab, August 23-29, 1979, ed. by
T. B. W. Kirk and H. D. I. Abarbanel (Fermi National Accelerator Laboratory,
Batavia, IL, 1979}
\def \hb87{{\it Proceeding of the 1987 International Symposium on Lepton and
Photon Interactions at High Energies,} Hamburg, 1987, ed. by W. Bartel
and R. R\"uckl (Nucl.\ Phys.\ B, Proc.\ Suppl., vol.\ 3) (North-Holland,
Amsterdam, 1988)}
\def \ib{{\it ibid.}~}
\def \ibj#1#2#3{~{\bf#1}, #2 (#3)}
\def \ichep72{{\it Proceedings of the XVI International Conference on High
Energy Physics}, Chicago and Batavia, Illinois, Sept. 6 -- 13, 1972,
edited by J. D. Jackson, A. Roberts, and R. Donaldson (Fermilab, Batavia,
IL, 1972)}
\def \ijmpa#1#2#3{Int.\ J.\ Mod.\ Phys.\ A {\bf#1}, #2 (#3)}
\def \ite{{\it et al.}}
\def \jhep#1#2#3{JHEP {\bf#1}, #2 (#3)}
\def \jpb#1#2#3{J.\ Phys.\ B {\bf#1}, #2 (#3)}
\def \lg{{\it Proceedings of the XIXth International Symposium on
Lepton and Photon Interactions,} Stanford, California, August 9--14 1999,
edited by J. Jaros and M. Peskin (World Scientific, Singapore, 2000)}
\def \lkl87{{\it Selected Topics in Electroweak Interactions} (Proceedings of
the Second Lake Louise Institute on New Frontiers in Particle Physics, 15 --
21 February, 1987), edited by J. M. Cameron \ite~(World Scientific, Singapore,
1987)}
\def \kdvs#1#2#3{{Kong.\ Danske Vid.\ Selsk., Matt-fys.\ Medd.} {\bf #1},
No.\ #2 (#3)}
\def \ky85{{\it Proceedings of the International Symposium on Lepton and
Photon Interactions at High Energy,} Kyoto, Aug.~19-24, 1985, edited by M.
Konuma and K. Takahashi (Kyoto Univ., Kyoto, 1985)}
\def \mpla#1#2#3{Mod.\ Phys.\ Lett.\ A {\bf#1}, #2 (#3)}
\def \nat#1#2#3{Nature {\bf#1}, #2 (#3)}
\def \nc#1#2#3{Nuovo Cim.\ {\bf#1}, #2 (#3)}
\def \nima#1#2#3{Nucl.\ Instr.\ Meth. A {\bf#1}, #2 (#3)}
\def \npb#1#2#3{Nucl.\ Phys.\ B~{\bf#1}, #2 (#3)}
\def \os{XXX International Conference on High Energy Physics, 27 July
-- 2 August 2000, Osaka, Japan}
\def \PDG{Particle Data Group, D. E. Groom \ite, \epjc{15}{1}{2000}}
\def \pisma#1#2#3#4{Pis'ma Zh.\ Eksp.\ Teor.\ Fiz.\ {\bf#1}, #2 (#3) [JETP
Lett.\ {\bf#1}, #4 (#3)]}
\def \pl#1#2#3{Phys.\ Lett.\ {\bf#1}, #2 (#3)}
\def \pla#1#2#3{Phys.\ Lett.\ A {\bf#1}, #2 (#3)}
\def \plb#1#2#3{Phys.\ Lett.\ B {\bf#1}, #2 (#3)}
\def \pr#1#2#3{Phys.\ Rev.\ {\bf#1}, #2 (#3)}
\def \prc#1#2#3{Phys.\ Rev.\ C {\bf#1}, #2 (#3)}
\def \prd#1#2#3{Phys.\ Rev.\ D {\bf#1}, #2 (#3)}
\def \prl#1#2#3{Phys.\ Rev.\ Lett.\ {\bf#1}, #2 (#3)}
\def \prp#1#2#3{Phys.\ Rep.\ {\bf#1}, #2 (#3)}
\def \ptp#1#2#3{Prog.\ Theor.\ Phys.\ {\bf#1}, #2 (#3)}
\def \rmp#1#2#3{Rev.\ Mod.\ Phys.\ {\bf#1}, #2 (#3)}
\def \rp#1{~~~~~\ldots\ldots{\rm rp~}{#1}~~~~~}
\def \si90{25th International Conference on High Energy Physics, Singapore,
Aug. 2-8, 1990}
\def \slc87{{\it Proceedings of the Salt Lake City Meeting} (Division of
Particles and Fields, American Physical Society, Salt Lake City, Utah, 1987),
ed. by C. DeTar and J. S. Ball (World Scientific, Singapore, 1987)}
\def \slac89{{\it Proceedings of the XIVth International Symposium on
Lepton and Photon Interactions,} Stanford, California, 1989, edited by M.
Riordan (World Scientific, Singapore, 1990)}
\def \smass82{{\it Proceedings of the 1982 DPF Summer Study on Elementary
Particle Physics and Future Facilities}, Snowmass, Colorado, edited by R.
Donaldson, R. Gustafson, and F. Paige (World Scientific, Singapore, 1982)}
\def \smass90{{\it Research Directions for the Decade} (Proceedings of the
1990 Summer Study on High Energy Physics, June 25--July 13, Snowmass,
Colorado),
edited by E. L. Berger (World Scientific, Singapore, 1992)}
\def \tasi{{\it Testing the Standard Model} (Proceedings of the 1990
Theoretical Advanced Study Institute in Elementary Particle Physics, Boulder,
Colorado, 3--27 June, 1990), edited by M. Cveti\v{c} and P. Langacker
(World Scientific, Singapore, 1991)}
\def \yaf#1#2#3#4{Yad.\ Fiz.\ {\bf#1}, #2 (#3) [Sov.\ J.\ Nucl.\ Phys.\
{\bf #1}, #4 (#3)]}
\def \zhetf#1#2#3#4#5#6{Zh.\ Eksp.\ Teor.\ Fiz.\ {\bf #1}, #2 (#3) [Sov.\
Phys.\ - JETP {\bf #4}, #5 (#6)]}
\def \zpc#1#2#3{Zeit.\ Phys.\ C {\bf#1}, #2 (#3)}
\def \zpd#1#2#3{Zeit.\ Phys.\ D {\bf#1}, #2 (#3)}

\end{document}